\documentclass{nature}

\usepackage{amsmath}
\usepackage{amssymb}
\usepackage{graphicx}
\usepackage{upgreek}
\usepackage{float}
\usepackage{bigints}

\title{The Future of Fast Radio Burst Science}

\author{E. F. Keane$^{1}$}

\begin{document}

\maketitle

\begin{affiliations}
 \item SKA Organisation, Jodrell Bank Observatory, SK11 9DL, UK. 
\end{affiliations}


\newpage
\begin{abstract}
The field of Fast Radio Burst (FRB) science is currently thriving and growing rapidly. The lines of active investigation include theoretical and observational aspects of these enigmatic millisecond radio signals. These pursuits are for the most part intertwined so that each keeps the other in check, characteristic of the healthy state of the field. The immediate future for FRB science is full of promise --- we will in the next few years see two orders of magnitude more FRBs discovered by the now diverse group of instruments spread across the globe involved in these efforts. This increased crop, and the increased information obtained per event, will allow a number of fundamental questions to be answered, and FRBs' potential as astrophysical and cosmological tools to be exploited. Questions as to the exact detailed nature of FRB progenitors and whether or not there are one or more types of progenitor will be answered. Questions as to source counts, the luminosity distribution and cosmological density of FRBs will also be addressed. Looking further ahead, applications involving FRBs at the highest redshifts look set to be a major focus of the field. The potential exists to evolve to a point where statistically robust cosmological tests, orthogonal to those already undertaken in other ways, will be achieved. Related work into FRB foregrounds, as well as how to identify new events in ever more challenging radio-frequency interference environments, also appear likely avenues for extensive investigations in the coming years.
\end{abstract}

Gaining a snap-shot summary of the current state of FRB science is no easy task as the field continues to develop at an astounding pace. Including the various 
yet-to-be published events,
most FRBs have been discovered since writing began for this article (July 2018)! Giving an outlook on the future of FRB science is consequentially even more difficult and involves a rather large extrapolation into the coming years. However it is surely more tractable a proposition than if the founder of the field, Duncan Lorimer, had been asked to predict the current state of affairs back in 2007. Personally speaking, it is extremely satisfying to see the trajectory of FRB science in recent times and especially since 2013; in the early years it was a somewhat lonely field to work in at times.
This all changed as the rate of discovery picked up, but until quite recently it is fair to say that FRB science has been an immature field of research. From 2007 to the time of writing it has been the case that $N_{\mathrm{theory}}/N_{\mathrm{FRB}} > 1$, where $N_{\mathrm{theory}}$ is the number of theories for what FRBs are and $N_{\mathrm{FRB}}$ is the number of FRBs known. There have been time intervals when the time-derivative of this maturity metric far exceeded unity. However this now seems to have changed, and FRB science looks to have come of age by these and many other measures that are outlined below. 
Now, and in the future, the two over-arching questions that must be addressed by FRB science are: (1) what are FRBs?; (2) what are FRBs good for? The now rather large FRB community can be sub-divided into approximatley three groups according to how much they are focusing on one, the other or both of these big-picture questions. Below we briefly address these two questions before looking to the near- and longer-term future of FRB science when these, and many other more detailed questions, will be addressed.

\textit{What Are They?} This question is addressed in detail elsewhere\cite{letter,pen} but here we give a first-order answer to this fundamental query. FRBs are characterised by large dispersion measures (DM), the integrated electron density along the line of sight to the source, with values exceeding the maximum Galactic contribution from the interstellar medium (ISM) by as much as\cite{cl02,ymw16,superb2} a factor of $200$. To explain such large excess DM values necessitates a large contribution from the intergalactic medium (IGM). As the electron density of the IGM is typically $\sim 10^{-5}$ that of the ISM\cite{inoue04} one must immediately infer distances of several gigaparsecs and non-negligible redshifts, $z$. Host contributions (from progenitor and/or host galaxy) could account for some of the DM, but large contributions might be thought to be disfavoured geometrically\cite{xh15} and in any event are suppressed by a factor $(1+z)$ in our frame of rest\cite{mkg+15}. Thus it is difficult not to conclude\cite{me18b} that the majority of the DM for the majority of FRBs comes from the IGM, which is perhaps the most interesting point about FRBs. If at such distances, but still detected as jansky-level sources with radio telescopes on Earth they must then have high radio luminosities. Feeding this into a simple calculation of brightness temperature implies that the emission is necessarily non-thermal with a high coherence factor (see Figure 1). From our knowledge of pulsar emission, in particular the Crab pulsar's so-called `giant pulses' which have comparable brightness temperatures\cite{hej16}, we might then conjecture that there are strong magnetic fields at play and that the emitted radiation may be highly polarised. Basic causality arguments tell us that, from the observed time duration of the bursts, the physical distance scales of relevance are kilometres --- in the most recent FRB discoveries features on $\sim 10$-$\upmu$s time scales have been observed\cite{ffb+18} --- implying that FRB progenitors likely involve compact objects, and in particular neutron stars. The above deductions were mostly obvious from the first FRB detection\cite{lbm+07}, but initial focus was quite rightly put onto confirming their astrophysical nature\cite{bbe+11,kskl12,tsb+13,sch+14,pkb+15,cfb+17}. One of many things learnt from that work was the confidence that FRBs are \textit{bona fide} astrophysical sources, and furthermore that they are numerous. Lower-limit estimates for the number of these events occurring are a few thousand each day \cite{superb2}.

\textit{What Are They Good For?} A large number of applications of FRBs have been identified in the literature\cite{jp}, of which we now briefly mention some illustrative examples.
It is often said that FRBs can be used to `weigh the Universe', but to be more accurate one can use FRBs to weigh the electrons associated with the ionised component of the baryons in the IGM. The dispersion of an FRB signal is caused by all of the ionised baryons along the line of sight --- this includes the so-called ``missing baryons''\cite{breg07,shull12} that have not been directly observed but are inferred to exist from the standard cosmological model and space-based CMB observations\cite{wmap13}. To identify these missing baryons requires measuring a sufficiently large sample of $\sim 100$ localised FRBs with independently measured redshifts. By quantifying the average DM as a function of $z$ one can measure cosmological parameters such as $\Omega_{\mathrm{baryon}}$, the energy density of baryons. 
Measuring the distribution of DM as a function of $z$ clearly requires a larger sample, but in doing so one could determine where the baryonic content is located, e.g. whether it is within or well beyond galaxy halos\cite{mcq14}.
The high-z applications of DM-based tests are perhaps the most powerful, and are considered in more detail below. 
But beyond these one can use the polarisation properties of FRB signals to study magnetic fields in the IGM and/or the host environments. The data do show evidence for strong magnetic fields and high polarisation fractions\cite{mls+15,msh+18} but the picture is far from uniform across the sub-sample where this information exists\cite{superb3}.
The total intensity time-frequency profiles of FRBs can be used to study IGM turbulence\cite{rsb+16}. By quantifying the effects of multi-path propagation (usually termed `scattering') and how this broadens the FRB pulses in a frequency-dependent way, one can determine the length scales of IGM turbulence with very fine spatial resolution of perhaps pico-arcseconds\cite{mk13}. With a sufficiently large sample one could perform tomography of the cosmic web. 

At the time of writing there are six telescopes which have discovered an FRB and many teams using these and other facilities are searching the skies for more. Figure 2 shows the FRBs which have been detected as a function of time up to and including the first half of 2018. It all started with Parkes; whereas the first FRB we now know about hit `The Dish' in June 2001 (FRB~010621)\cite{kkl+11,kskl12}, it was one that did so in July 2001 (FRB~010724) that was the first to be noticed in the data\cite{lbm+07}, albeit not until 2007. In most of the interval between its occurrence and detection there was a reduced emphasis on high time resolution (i.e. pulsar) surveys at Parkes after the conclusion of the Parkes Multi-beam Pulsar Survey (1997--2002)\cite{mlc+01}. Those that did occur in this time\cite{bjd+06,jbo+09,bkl+13} were limited in observing time and sky coverage, and as a result there is a complete absence of FRBs detected between 2002 and 2008 inclusive. However, a year after the discovery of FRB~010724, the High Time Resolution Universe (HTRU) surveys\cite{kjs+10,bck+13} begun with an extensive high Galactic latitude component that was designed specifically to find more such events. Ten HTRU FRB discoveries\cite{tsb+13,cha+16} followed quickly on the heels of two archival discoveries both from 2001 data\cite{kkl+11,bb14}. As of July 2018, Parkes has discovered 27 FRBs. Of these, 23 have been in the past $\sim7$ years, in which time $\sim 1$ year has been unavailable due to commissioning new receivers, so that the long-term FRB rate at Parkes has been $\sim 4$ FRBs/year for the typically committed observing time. 

Arecibo joined the fray with the discovery of FRB~121102 (hereafter `the repeater')\cite{sch+14} which has subsequently been observed to show hundreds of repeat bursts\cite{ssh+16,hds+17,ska_france,sbh+17,priv_comm_stappers}. This is in stark contrast to all other FRBs known to date which have not repeated, despite much longer times on sky in search of repetition. The distribution of wait times between bursts from the repeater do not seem to be random, not following an exponential distribution; instead they are seen to be clustered in time and can be modelled by a Weibull distribution\cite{oyp18}. Despite this, with a reliably repeating source it is a matter of time and patience before one can localise the source and this was achieved to $\sim 0.1$~arcsec resolution\cite{clw+17}. The source was shown to lie in a region of ongoing star formation within a $z = 0.19273(8)$ low-metallicity dwarf galaxy\cite{tbc+17,bta+17}, again consistent with the idea that FRBs originate from neutron stars. Whereas it is generally possible to identify underlying rotation periods of intermittently emitting pulsars if one, as a rule of thumb, has $8$ events within an hour\cite{thesis}, this has not proven possible with the repeater. In fact, the lowest common denominator of the time differences between pulses is approximately the duration of a single pulse\cite{hds+17,katz18}. This might mean that it is impossible to resolve any underlying pulse period. Furthermore, some models for the progenitor propose a flaring magnetar that is just a few tens-of-years old\cite{bel17} --- in such a scenario the slow-down rate is expected to be significant. As modelling the latter requires independent period estimates at many epochs; the prospects for identifying this, if this were the underlying scenario, are close to zero. Very recently, the Arecibo team has identified a second FRB that is rather faint\cite{pat18}; it is unclear as yet if this source repeats.

The upgraded Molonglo Observatory Synthesis Telescope (UTMOST)\cite{bjf+17} has been operating an FRB search since 2016 and has so far discovered 6 FRBs. 
Green Bank has discovered 1 FRB in archival data\cite{mls+15}, but the Green Bank North Celestial Cap (GBNCC) survey, operating at $350$~MHz, has not as yet detected an FRB\cite{ckj+17}. 
On a more industrial scale, and operating in a fly's eye mode with a wide field-of-view but only single 12-m dish sensitivity\cite{bsm+17,askap} ASKAP has found 26 FRBs as of mid-2018 using a sub-array with the number of dishes, $N_{\mathrm{dish}}$, varying between 6 and 10. With this impressive demonstration of capability complete, it has recently changed focus, pointing all dishes towards the same sky location in an effort at precise localisation. This operating method reduces the field-of-view, and hence the rate, by a factor $N_{\mathrm{dish}}$, but with an incoherently combined array the rate is also improved by a factor $N_{\mathrm{dish}}^{\alpha/2}$, where $\alpha$ is the uncertain source-count distribution slope\cite{me18b}, which ASKAP will help to determine.
The Canadian Hydrogen Intensity Mapping Experiment (CHIME) has just reported its first FRB discovery in July 2018\cite{chime} and will likely leap-frog the other instruments in terms of the number of FRBs detected. The CHIME detections, apparently extending down to frequencies as low as $400$~MHz, also imply that the GBNCC\cite{ckj+17} and other surveys in this band may yet yield some FRBs as questions as to scatter broadening or even emission of FRBs in this band seem to have been addressed. 

There are a number of open questions in FRB science at present --- some are currently being tackled, others will shortly see progress, and yet more will not be addressed for a while yet --- here we look at some of the most pertinent ones. i) \textit{How many FRBs are there?} This extremely basic question is quite important for a number of reasons. Firstly, it gives a first order idea of what the progenitors may or may not be, i.e. if the rates match or deviate from those of known events like supernovae, gamma-ray bursts, blitzars\cite{fr14}, etc. Secondly, from the point of view of discovering more FRBs with new telescopes/receivers/pipelines one needs to know the observed rates for a reference set up (this has thus far been the current Parkes configuration\cite{swb+96,psb+16}) so as to perform various scalings based on an understanding of the parameter space searched. Perhaps surprisingly it is rather difficult to answer this question, as it relies on a knowledge of telescope calibration at a level that has not been determined. In the case of Parkes one does not have a full understanding of the multi-beam receiver sky response to the required level, particularly off-axis beyond the various half-power surfaces. The uncertain location in the beam results in an uncertain brightness, so that the most basic question of how many FRBs brighter than some value occur is uncertain\cite{me18a}. The more difficult question of converting to a volumetric rate is more uncertain. Pipeline efficiences (see below) and changing radio-frequency interference (RFI) environments\cite{pkb+15,superb1}, which are often not well monitored, also change the effective thresholds. While it is reasonably straight forward to work out the rate for the Parkes set-up in FRBs per unit time, in converting to meaningful physical units the best estimates are lower limits evading the above problems in the simplest way possible. Much progress could potentially be made by taking the time to carefully perform the necessary calibration measurements for all of the telescopes discovering FRBs, something that is intended with new receivers. In a typical symptom of a fast-moving field these simple but laborious measurements have not yet been made. As any two FRBs with the same intrinsic brightness are not equally detectable in a way that depends on the properties of their lines of sight\cite{kp15}, i.e. they have the same fluence\footnote{It is standard practice to refer to the band-averaged specific fluence simply as `fluence'.}, but will have different detected flux densities, fluence-complete lower limit rates discarding those in an incompleteness region are perhaps the most useful for simple scalings. The rate of FRBs, as determined from the Parkes sample\cite{superb2}, is $>1.7^{+1.5}_{-0.9}\times 10^3$~FRBs$/(4\pi$~sr)/day, above a fluence of $\sim 2$~Jy\,ms. In addition to Parkes, the only other instruments with a reasonable sample now, or soon, are ASKAP and CHIME and these also will have to characterise their telescope response and the environmental conditions, and the stability thereof, to quite a precise level to enable meaningful rate estimates. 

ii) \textit{What is the $\log N-\log S$ distribution?} Another basic observable is the brightness distribution, $\log N-\log S$, where $S$ is flux density, or more meaningfully (see above) $\log N-\log F$, where $F$ is fluence. In a Euclidean Universe one expects $N\propto F^{-3/2}$, but for a cosmological population in an evolving Universe one expects deviations from that\cite{rc61}. While it seems clear from other indicators that the population is cosmological (i.e. the range of DM values requiring a majority IGM component, the localisation of the repeater, the Galactic latitude distribution\cite{superb2}) this basic test cannot yet be performed. If one discards those FRBs in the fluence-incomplete region of parameter space, one is left with an insufficiently small sample. One can simply observe that if one bins the remaining FRBs in (say) decadel fluence bins, then addressing the question as to whether the distribution is Euclidean requires far in excess of $\sim (10)^{3/2}$, $\sim 32$ FRBs, above the completeness limit to see the effect and to beat down shot noise. As it is, the slope of the distribution is highly sensitive to single events\cite{vrhs16}, i.e. it fails a basic boot-strapping test. The best current estimate\cite{me18a} for the power-law slope is $\alpha = -2.6_{-1.3}^{+0.7}$, but debate over how to treat FRB~010724 considering the ``Winner's curse''\cite{me18a} and the use of optimised versus detection signal-to-noise ratio values are ongoing. 

iii) \textit{What is the pulse width distribution?} The observed and intrinsic temporal durations of FRBs are usually referred to as their `widths' in time. The observed distribution sees many FRBs that are clearly resolved in time at several milliseconds, some of these with multiple components\cite{cha+16}. Also, there are many FRBs that are unresolved in time $\lesssim 0.5$~ms, whose duration is limited by dispersion smearing due to finite frequency resolution\cite{kjb+16}. Overall the spread is an order of magnitude when considering pulse widths averaged over several hundred MHz bands. On the unresolved end of this range the intrinsic pulse widths could be much narrower. The broad-band nature of FRBs seems to show emission bands of a few hundred MHz at frequencies up to $\sim 1.6$ GHz, or $\sim 8$~GHz in the case of the repeater\cite{gsp+18}. Such bandwidths suggest intrinsic timescales possibly as narrow as $\sim 10$~ns. Where finer frequency resolution has been possible narrow band structures, down to $\sim 10\;\upmu$s scales, have been seen\cite{ffb+18}. The distribution of pulse widths that has been observed are often ignored, but as different progenitor models predict different timescale FRBs it may offer a means of distinguishing these if there are two or more progenitor populations.

iv) \textit{How many missing FRBs are there in the existing data?} There are likely quite a number of FRBs already recorded in the existing archives of survey data which have not yet been noticed. The most basic reason for this is that some of the archival data has never been searched. For those that have been searched it has often not been thorough or complete and fundamental pipeline issues have been identified which result in needless degradations in signal-to-noise ratios, and consequentially reduced search volumes and missed FRBs\cite{kp15}. For the most part pipelines are not tested well or even at all, beyond binary tests such as whether or not they can detect usually quite strong test signals. Full injection studies where pipelines are tested end-to-end\cite{pat18} for accuracy and reproducibility with meaningful test vectors are almost non-existant. It would seem incredible that this could remain the state of affairs considering the possible gains: one can obtain results equivalent to that of a much larger telescope by simply improving data processing pipelines. Another problem, which is more difficult algorithmically, is how to deal with RFI whose characteristics change in time, frequency, location and strength on many scales. A robust dynamic solution to this ever worsening issue is badly needed for radio astronomy in general. Another limitation arises from limited data rates --- if data could be recorded more quickly then the possibility of detecting narrower FRBs is opened. At present even very high flux density FRBs with short durations can be undetectable due to this effect. One final problem which may result in missed FRBs stems from how to define an FRB in a basic observational sense. Any definition must be in reference to a model for the Milky Way's electron density model. The two current foremost models\cite{cl02,ymw16}, NE2001 and YMW16, are both uncertain and disagree to different magnitudes for different lines of sight meaning there are many `grey areas' where one is uncertain whether a given source is Galactic or not\cite{keane16}. The upshot is that some sources are FRBs according to YMW16, but not according to NE2001, with the opposite possible for some lines of sight\cite{rl17}. The FRBs that would be missed due to this confusion are very nearby and so of no use for cosmological tests but for studies of FRB progenitors are of the highest interest.
Questions as to the rates, source count distribution, pulse width distribution and completeness of surveys are all entwined, and must each be very carefully treated. Ignoring any of these aspects can result in wildly varying and incorrect predictions for the others.

v) \textit{What are the polarisation properties of FRBs?} At the time of writing eight FRBs had polarisation properties reported in the literature\cite{pbb+15,mls+15,kjb+16,rsb+16,pbk+17,msh+18,superb3}. Amongst this group there is no consensus as to what an FRB looks like in terms of its polarised light, and the characteristics of each signal are almost unique. Two sources (FRBs~140514 and 160102) show appreciable circular polarisation, whereas most show none or are consistent with zero within the limits of the signal-to-noise ratio. Six sources (FRBs~110523, 121102, 150215, 150418, 150807 and 160102) show linear polarisation allowing a measurement/limit of the rotation measure (RM) --- values consistent with zero and/or the Galactic contribution have been seen in three cases. In two other cases RM values an order of magnitude in excess of the Galactic contribution have been seen, with FRB~160102 having an observed $\mathrm{RM}=-221(6)\;\mathrm{rad}\,\mathrm{m}^{-2}$, which is $\sim -2400\;\mathrm{rad}\,\mathrm{m}^{-2}$ when corrected for the Milky Way contribution and red-shifted to the frame of emission for its nominal DM-inferred redshift. This however is dwarfed by the repeater which has $\mathrm{RM}\approx 10^{5}\;\mathrm{rad}\,\mathrm{m}^{-2}$, which is four orders of magnitude above the Galactic contribution. From this disparate group of properties one can not conclude much, other than to say that the allowed DM and RM ranges for various models imply there is no one-solution-fits-all model\cite{csp16,superb3}. The majority of reported FRBs have only total intensity Stokes $I$ information, but this will soon change, and there are at least four more likely to be reported imminently\cite{pgd+18,osj+18a,osj+18b,osj+18c}.

vi) \textit{Do all FRBs repeat?} Statistically speaking, it is essentially impossible that FRB~121102 is the only source of its kind in the Universe, but whether or not the other FRBs now known show repeat bursts is a different question and the subject of much active research\cite{letter}. Apart from discovering more repeaters, for which there are strategies\cite{fll18}, determining how typical or not the repeater is can only be answered by discovering more `one-off' events, and limiting their repetition rates as much as is feasible. While it is impossible to prove something will never repeat one can reduce the allowed parameter space beyond the point of credibility. While repetition speaks to the details of the progenitor system being non-catastrophic, for the purposes of using FRB signals to probe aspects of the Universe, it would seriously improve matters. Not only would all FRBs be localised rather easily (with redshifts following) if repetition was a given, one could get ever improving accuracy on DM, RM, polarisation and scattering parameters to feed into some of the future investigations discussed below.

The future of FRB science will involve a sample increased by orders of magnitude, with an appreciable fraction that are localised and spanning a wider range of redshifts than at present. The focus will thus naturally be on high-redshift measurements. These can be used to measure cosmological parameters such as the energy density of matter ($\Omega_{\mathrm{m}}$), baryons ($\Omega_{\mathrm{baryon}}$), dark energy ($\Omega_{\Lambda}$), curvature ($\Omega_{\mathrm{k}}$), the ionisation fraction profile of the IGM ($f_{\mathrm{IGM}}(z)$), the dark energy equation of state parameter ($\omega(z)$) and the reionisation histories for Helium ($X_{\mathrm{He}}(z)$) and Hydrogen ($X_{\mathrm{H}}(z)$). Each of these properties results in a different characteristic feature in the $\langle \mathrm{DM}(z) \rangle$ relation, the \textit{average} DM as a function of redshift, which is given by:
\begin{equation}
  \langle \mathrm{DM}(z) \rangle = \bigint_{0}^{z} \frac{f_{\mathrm{IGM}}(z')(1+z')\left(\frac{3}{4}X_{\mathrm{H}}(z')+\frac{1}{8}X_{\mathrm{He}}(z')\right)dz'}{\left( \Omega_{\mathrm{m}}(1+z')^3 + \Omega_{\mathrm{k}}(1+z')^2 + \Omega_{\Lambda}\exp\left( 3 \int_{0}^{z'} \frac{(1+\omega(z''))dz''}{1+z''} \right) \right)^{1/2}} \;.
\end{equation}
Equation 1 is shown in Figure 3 for standard cosmological parameters with two illustrative reionisation histories. It is important to mention that none of the FRBs so-far discovered can be used for these tests, due to poor sky localisation. To measure the above quantities requires a sufficiently large sample of FRBs with both DM measurements and redshifts, the latter measured in optical/infrared observations of the host galaxies. Below we look at projections on future FRB discoveries in terms of yields and localisation; it seems clear that a sufficient sample will be forthcoming. While there are line-of-sight variations\cite{ioka03} the average DM, $\langle \mathrm{DM}(z) \rangle$ is a well-defined quantity. Next generation tests will examine the distribution of DM values as a function of $z$, but first generation tests will focus on $\langle \mathrm{DM}(z) \rangle$ measurements, which require approximately an order of magnitude less events to realise. The first measurement might be the ``missing baryons'', which can be performed using all FRBs below the Helium Epoch of Reionisation (EoR) which is thought to occur\cite{sah02} in the range $3\lesssim z \lesssim 4$. This will yield $f_{\mathrm{IGM}}(z)\Omega_{\mathrm{baryon}}$; $f_{\mathrm{IGM}}(z)$ is quite poorly constrained with the best estimate\cite{zlw+14} being a value of $0.82$ at $z=0$ linearly increasing to $0.90$ at $z=1.5$ and staying steady at this value for higher redshifts. We note that the $\langle \mathrm{DM}(z) \rangle$ relation scales up or down linearly with $f_{\mathrm{IGM}}(z)$, e.g. the commonly used approximation\cite{frbcat} of $\langle \mathrm{DM}(z) \rangle = 1200z$ is too high as it arises from assuming $f_{IGM}(z)=1$, as well as an additional $(8/7)$ factor from ignoring Helium reionisation, from early calculations that were not intended to be accurate to this level\cite{inoue04}.

At $z\lesssim 0.5$ the Milky Way and host DM contributions are relatively large, and are poorly constrained so could be problematic for the $f_{\mathrm{IGM}}(z)\Omega_{\mathrm{baryon}}$ measurement. As most FRBs will be fainter (they do not seem to be standard candles\cite{askap,dunc}) this might be a large sample. Without proper FRB foreground information it might be required to ignore the lowest redshift sub-sample in the assessement of $\langle \mathrm{DM}(z) \rangle$. At present, even if all FRBs known were localised one might have to discard all with $\mathrm{DM}\lesssim 500\;\mathrm{cm}^{-3}\;\mathrm{pc}$ which is far from ideal. This highlights the need for a large scale observational project to reduce uncertainties in the electron density model for the Galaxy. Foreground work must also include the often-ignored contributions from the Galactic halo\cite{dgbb15} which are typically $\sim 30\;\mathrm{cm}^{-3}\;\mathrm{pc}$. Knowledge of this component could be limiting even for high redshift tests, e.g. whether $\omega$ has a value of $-1.05$, $-1.00$ or $-0.95$ imparts a deviation of $\sim 30\;\mathrm{cm}^{-3}\;\mathrm{pc}$ to the observed $\langle \mathrm{DM}(z) \rangle$.

Moving to higher redshifts around the Helium EoR, the exact EoR profile results in differences of $\sim 100\;\mathrm{cm}^{-3}\;\mathrm{pc}$ in the observed $\langle \mathrm{DM}(z) \rangle$ relation. This should be readily measurable\cite{zok+14}, and the prospects of detecting FRBs to this redshift are good. FRB progenitors are likely neutron stars and so their cosmological density should map the star formation rate which peaks\cite{md14} in the range $2\lesssim z \lesssim 3$, so that by $z\approx 4$ there should still be sufficient FRB sources. Observationally, our current highest gain instruments with appeciable fields of view (MeerKAT and Parkes) can detect the current FRB population scaled to $z=4$ (see Figure 1) with one example in particular (FRB~160102) quite likely to be beyond $z=2$ for most scenarios. Detecting FRBs to the Hydrogen EoR is considerably more uncertain and possibly unachievable for a number of reasons (primarily a lack of sources and sensitivity limits), but $\langle \mathrm{DM}(z) \rangle$ does hit a ceiling at the beginning of the Hydrogen EoR which would be clear if any sources were ever detected. The Hydrogen EoR ceiling is at $\sim 5000\;\mathrm{cm}^{-3}\;\mathrm{pc}$ ($\sim 6000\;\mathrm{cm}^{-3}\;\mathrm{pc}$) for a step-function EoR profile at $z=7$ ($z=9$), DM values which are now trivial to search to. This was not always the case, with maximum DM values limited by computing (or in some cases by arbitrary cut-offs), which resulted in FRBs being missed in earlier processings~\cite{lbm+07,kkl+11,bb14,superb2}. It is now standard practice to search to DM values of $10^{4}\;\mathrm{cm}^{-3}\;\mathrm{pc}$. While having highlighted numerous caveats we also note that many of the practical difficulties identified above do get better by simply adding ever more FRBs to the sample and perhaps taking advantage of such an abundance to ignore particularly poorly undersood (in terms of foregrounds) patches of sky. However the most precise tests at higher redshift, e.g. looking at prolonged EoR ionisation profiles, examining the redshift-dependence of $\omega$ etc. will ultimately be dominated by these foregrounds once $\sqrt{N_{\mathrm{FRB}}}$ errors are beaten down.

To fully address FRB science goals requires a large number of localised FRBs at a wide range of redshifts along well-understood lines of sight. The number of FRBs is not necessarily the best single metric and optimal FRB searches must consider other factors, e.g. arrays can choose between having a large yield of poorly localised more-nearby FRBs or a smaller crop of localised more-distant FRBs. Optimal FRB searches must consider the following parameters: observing time ($T_{\mathrm{obs}}$), field of view ($\Omega$), observing band ($\Delta f$), angular resolution ($\Delta\Omega$), sensitivity ($G$), time resolution ($\Delta t$) and reaction time ($t_{\mathrm{react}}$). 
The rate of FRB discoveries scales linearly with $T_{\mathrm{obs}}$ and $\Omega$. Maximising $T_{\mathrm{obs}}$ requires constantly searching for FRBs, which in practice means commensally `piggy-backing' all observations with FRB searches. This obvious, but perhaps difficult-to-realise, optimisation is achieved by the UTMOST telescope which searches for FRBs for all of the available observing time\cite{bjf+17}; this is similarly the case for CHIME\cite{chimefrb}, will shortly be the normal operating mode of MeerKAT as part of the MeerTRAP programme\cite{meertrap} and, when constructed, will likewise be the case for the full DSA array\cite{priv_comm_vikram}. At Parkes the FRB observing fraction is relatively high and at the tens of percent level, but plans are in preparation to commensally observe at all times, at least with the multi-beam receiver, meaning this will then be as large as it can possibly be for the instrument\cite{kg18}. The time spent on sky searching for FRBs by Green Bank, Effelsberg, Arecibo, and the Karl G. Jansky Very Large Array (VLA) is quite far from $100\%$ of the time so that upgraded capabilities along these lines remain an obvious step to boosting their FRB yields. If the Five-hundred metre Aperture Spherical Telescope (FAST) were to commensally search for FRBs at all times using its 19-beam receiver it too would be as optimised as possible\cite{fast}. Looking further ahead, the Square Kilometre Array (SKA) has designated FRBs as one of its 13 high priority science objectives for its initial deployment (SKA1) and is designed to commensally search for FRBs during pulsar searching, pulsar timing or imaging\cite{iau337}, meaning that it too will be optimised in terms of $T_{\mathrm{obs}}$. 
The field of view and observing band wherein to observe are related --- most FRBs have been discovered at $\sim 1.4$~GHz but the CHIME band between $400-800$~MHz should shortly leap-frog into first place. With all other quantities equal one would like to observe with the widest possible $\Omega$, which translates to the lowest observing frequency. Conversely searches above $\sim 2$~GHz are not likely to be as successful as $\Omega$ diminishes. It is unclear if the FRB search window will be shown to reach to even lower frequencies where one might take advantage of the vastly increased sky coverage --- efforts to date have been unsuccessful\cite{kca+15}. All of the afore-mentioned facilities have impressive $\Omega$ values with the exception of Arecibo, FAST, Green Bank and Effelsberg (until equipped with phased array feeds, PAFs, for the purpose); Parkes at $0.55\;\mathrm{deg}^2$ has the smallest $\Omega$ of the remainder and when this is boosted by a factor $\sim 3$ with the addition of a cooled PAF\cite{jimi_ursi} its capabilities will be maximised.
The ability to localise sources is crucial to the utility of the sample and this is where large single dishes lose out, and arrays excel. To maximise performance in this regard large single dishes (e.g. Arecibo, Effelsberg, FAST, Green Bank and Parkes) could be retro-fitted with cooled PAFs to maximise their FRB capabilities both in terms of yield and localisation. Efforts to these ends are in progress\cite{rss+18} but ultimately arrays dominate, and can always be extended both in number of elements and beamformer capability. Arrays can achieve $\sim 0.1$~arcsec resolution, but at the expense of survey speed for finite beam-forming capabilities, but this improves with time as computing hardware evolves, e.g. the survey speed of MeerKAT (at first operation) will be lower than Parkes\cite{iau337}, but should eventually be much better when the entire primary beam field of view is utilised. The ultimate combination to optimise $\Omega$ and $\Delta \Omega$ would be a large element array like the mid-frequency component of the first stage of SKA (SKA1-Mid), but where each dish is equipped with next generation PAFs. ASKAP and the Apertif project on Westerbork\cite{apertif}, which are smaller arrays but are already in operation, should shortly demonstrate the potential for PAFs in this domain. In terms of raw sensitivity, which maps to the maximum redshift to which an FRB would ever be detected, FAST edges SKA1-Mid, which in turn leads Arecibo, although in the case of the single dishes the rate of such events is severely compromised by the accessible field of view. Sensitivity has already been shown to be key for finding the most distant FRBs, needed for the high-$z$ cosmology tests --- the highest-DM FRB that has not been discovered by Parkes is\cite{cfb+17} FRB~160317 which has $\mathrm{DM} = 1111\;\mathrm{cm}^{-3}\,\mathrm{pc}$ --- despite the rather low rate of FRBs at Parkes it has discovered the top seven highest-DM FRBs.
One area where an abundance of hidden FRBs may come to light is at sub-millisecond time scales. The left-hand side of Figure 1, showing the fastest signals, is mostly empty due entirely to this region of parameter space not having been searched. The broadband nature of FRBs suggests this region may be populated and if it were possible in terms of data rates to get even slightly closer to raw (i.e. Nyquist) sampling of the data stream, such FRBs will be discovered. Parkes for example reduces its data rate by a factor of $25$ and this is typical\cite{kjs+10,psb+16,superb1}; this can reduce the signal-to-noise ratio of the fastest signals by up to a factor of $5$.

Amongst the telescopes that look set to have a low or zero FRB yield for the near future for the reasons outlined above are LOFAR, the VLA, Effelsberg, Arecibo and Green Bank. Well-localised\footnote{The localisation values represent those of one-off events, repeating FRBs will all be localised $\lesssim 1$~arcsec} (i.e. $\lesssim 1$~arcsec) FRBs numbering up to several tens per year will come imminently from both ASKAP and MeerKAT and, a few years later to $\sim 3$~arcsec with the DSA. UTMOST, when upgraded\cite{utmost2D} to UTMOST-2D, will find a few FRBs per year localised to $\sim 5$~arcsec levels. Parkes equipped with a cooled PAF can expect up to a few tens of FRBs per year localised to $\sim 30$~arcsec, Westerbork with the Apertif system will have a better yield but will only achieve similar localisation in one dimension as it is an East-West array\cite{apertif}. FAST's field of view is low, and its localisation will be $\sim 1$ arcmin, but its yield will depend on what $\log N- \log F$ is, but even if the number is small they may probe beyond the Helium EoR. CHIME will potentially find several hundreds of FRBs per year. Although these will only be localised to $\sim 30$~arcsec levels~\cite{chimefrb} its frequency range might allow localisation via HI absorption\cite{fo15} in some cases. SKA1-Mid (of which MeerKAT is a sub-array) is $\sim 5$ years behind the others in terms of getting on sky but will have a FAST-like sensitivity and offer a wide field of view and sub-arcsecond localisation. Together, this wide range of complementary instruments will shortly routinely identify hundreds of FRBs per year, with precise localisation of these gradually set to become the norm. The near-, mid- and long-term future of FRB science thus appears very bright. Characterising all of these discoveries will require quite a lot of 8- and 10-m optical/infrared telescope time so the most precisely localised sources will need to be prioritised. Additionally FRB scientists will continue to characterise their instruments and observatory environments better than ever before, and a similar approach will be needed for Galactic foregrounds in terms of the Milky Way's electron density distribution. Hopefully the achievements of FRB science during its second decade far surpass all the above predictions and extrapolations as much as they have done since 2007!


\begin{addendum}
\item The author would like to thank Professor Duncan Lorimer,
  Dr. Manisha Caleb, Dr. Jimi Green and the anonymous referee for
  helpful comments that improved the quality of this manuscript. The
  author would like to thank the ASKAP FRB team for providing advanced
  knowledge of their first 26 FRB discoveries, to the PALFA FRB team
  for advanced knowledge on FRB~141113, and to the DSA FRB team for
  advanced knowledge on the DSA specifications. 
 \item[Author Information] Reprints and permissions information is
   available at www.nature.com/reprints. The authors declare no
   competing financial interests. Readers are welcome to comment on
   the online version of the paper. Correspondence and requests for
   materials should be addressed to E.F.K. (E.Keane@skatelescope.org) \\
\end{addendum}

%
%


\newpage
\begin{figure}[H]
  \includegraphics[scale=0.6]{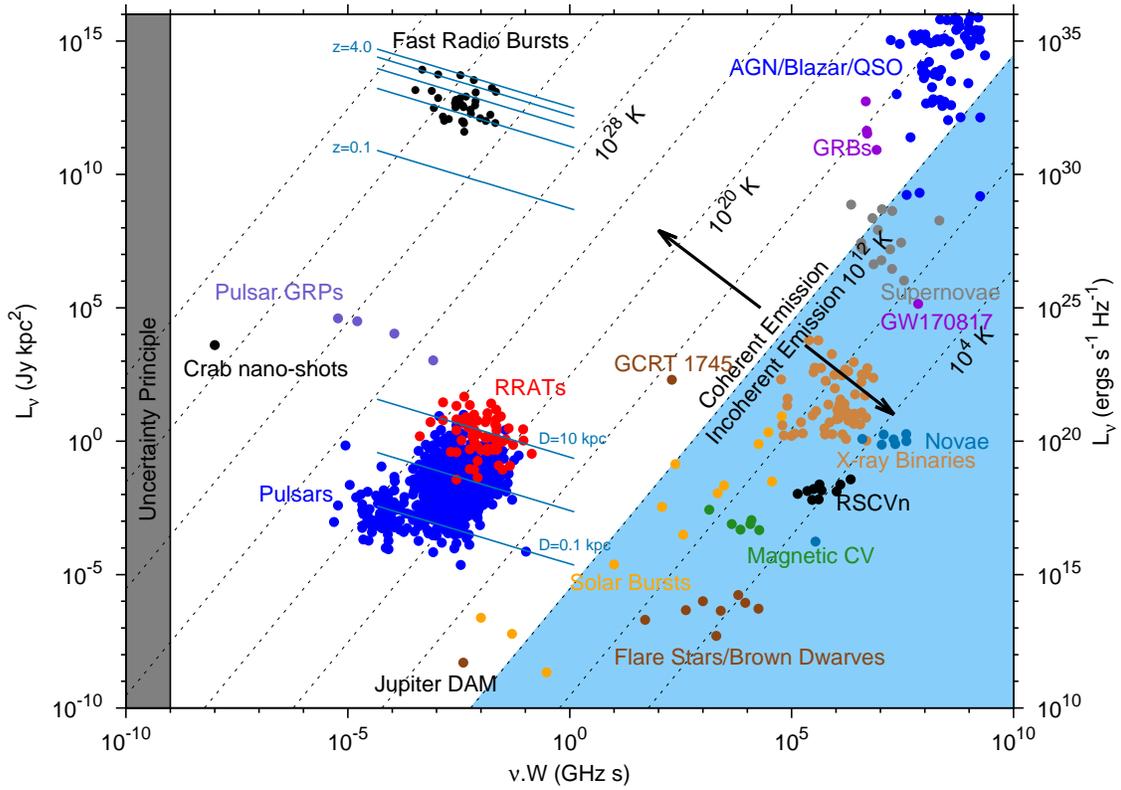}
  \caption{The transient parameter space showing radio luminosity on the vertical axis versus the product of observing frequency and timescale on the horizontal axis\cite{pfk15}. For illustrative purposes sensitivity curves for Galactic distances (0.1, 1 and 1~kpc) and various cosmological redshift values (0.1, 1, 2, 3 and 4) are shown that are appropriate for MeerKAT, or (to within the accuracy of the thickness of the lines) Parkes improved in gain by a factor of $3$ (indicative of an upgrade involving a cooled PAF).}
\end{figure}

\begin{figure}[H]
  \includegraphics[scale=0.5]{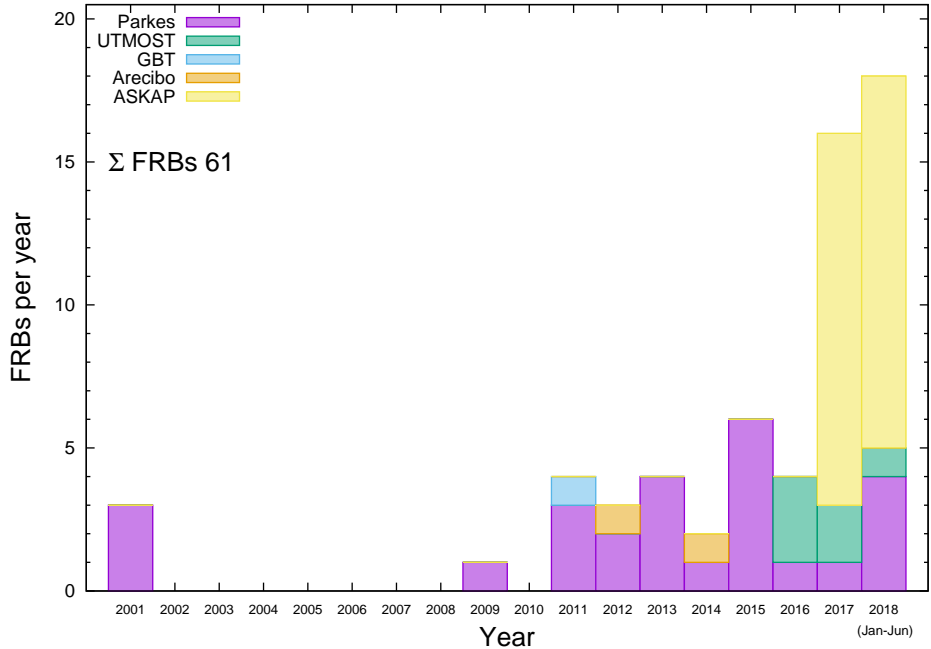}
  \caption{The number of FRBs detected each year since the first events in 2001 up to and including the first half of 2018. For the sake of having as complete a sample as possible up to as late a date as possible a cut-off has been placed in mid 2018, meaning one reported event each from Parkes\cite{osj+18c} and CHIME\cite{chime} are not displayed here. Until approximately 2016, only Parkes had a sufficient combination of time on sky, field of view, sensitivity and FRB search infrastructure to find an appreciable number of FRBs. Thus Parkes was for the most part the FRB work horse for the whole community, and in the time where it did not dedicate much time to high time resolution searches between 2002 and 2008 there is a clear dearth of detections. From approximately 2013 onwards the lag between FRB events occurring and being identified in the data reduced quite rapidly from years to seconds\cite{tsb+13,pbb+15,superb1}; this is thanks to progress in computing and in particular graphics processing units\cite{bbbf12}.}
\end{figure}

\begin{figure}[H]
  \includegraphics[scale=0.5]{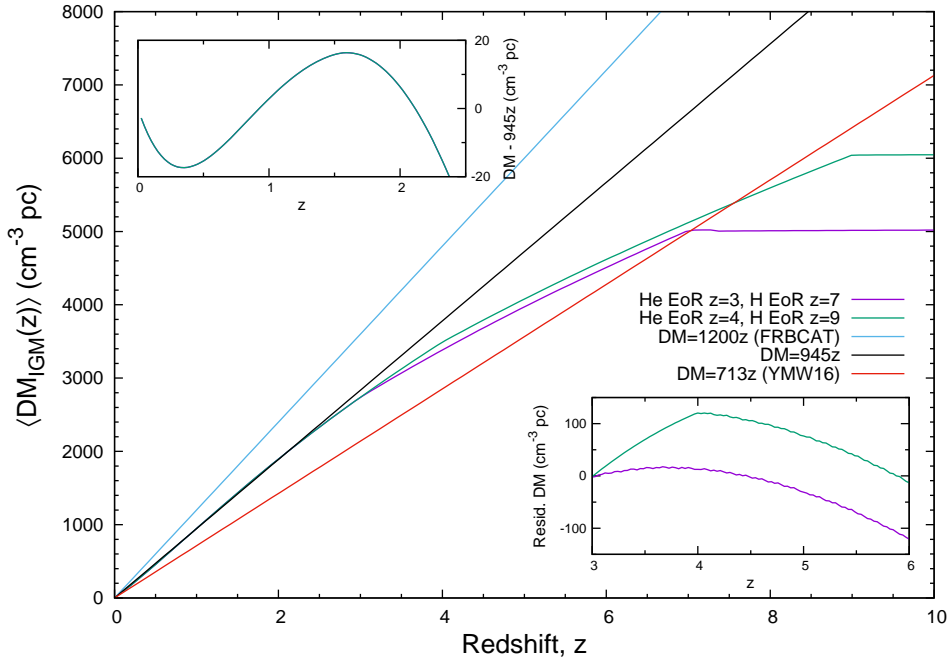}
  \caption{The main panel shows the average DM due to the IGM as a function of redshift (see main text). Equation 1 is plotted with standard cosmological parameters\cite{fp04,wmap13}, the best available estimate\cite{zlw+14} for $f_{\mathrm{IGM}}(z)$, and two realisations of step-function reionisiation events for Helium and Hydrogen at $z=3$ and $z=7$, and $z=4$ and $z=9$, respectively. Also over-plotted are two approximations that are in common use --- that of the FRB Catalogue\cite{frbcat} in blue and the YMW16 electron density model\cite{ymw16} in red. A third approximation is shown in black, and the top left inset panel shows the deviation of this third approximation as a function of redshift; it stays within $20\;\mathrm{cm}^{-3}\,\mathrm{pc}$ up to $z\approx 2.3$. The bottom right inset shows the $3<z<6$ region zoomed in, with the overall trend removed, highlighting the difference for the two different Helium reionisation redshifts considered; the difference between these two scenarios is $\sim 100\;\mathrm{cm}^{-3}\,\mathrm{pc}$.}
\end{figure}

\end{document}